\documentclass[runningheads,citeauthoryear]{apinv}
\usepackage{epsfig,cite,graphics}
\usepackage[T2A]{fontenc}
\usepackage[cp1251]{inputenc}
\usepackage{longtable}

\begin{document}

\title{Pre-main sequence stars}
\titlerunning{Pre-main sequence stars}
\author{Evgeni H. Semkov\inst{1}}
\authorrunning{E. Semkov}
\tocauthor{Evgeni Semkov}
\institute{Institute of Astronomy and National Astronomical Observatory, Bulgarian Academy of Sciences, Sofia, Bulgaria
        \newline
        \email{esemkov@astro.bas.bg}
}
\papertype{Dissertation summary. Accepted on 31.05.2023}
\maketitle

\section{Introduction}

The dissertation presents result from study of Pre-main sequence (PMS) stars that are in the earliest stages of stellar evolution. 
These young stellar objects are still in the process of formation, and the energy they emit is produced only by gravitational contraction.
The main results were obtained with the telescopes at the National Astronomical Observatory Rozhen, as well as with the use of archival photographic observations and spectral observations obtained in collaboration with colleagues from abroad. 
Our results are obtained by studying the photometric and spectral variability of PMS stars.
Our main goal is to study the processes of star formation, the formation of circumstellar disks, the structure of the circumstellar environment and the interaction of the star-disk system. 
The results of our research have been published in 65 scientific papers that have been cited over 300 times.

The photometric and spectral variability of PMS stars is of great importance for modeling star formation processes. 
On the one hand, photometric variability allows us to easily detect young objects, since they are characterized by rapid variability and in many cases with large amplitudes. 
On the other hand, stars form in groups and are physically located in the same geometric space, and several variable objects that are at approximately the same stage of evolution can be observed simultaneously. 
Comparing star systems of different ages can be used to trace the stages of stellar evolution.

\section{Photometric and spectral variability as a method to study physical processes in PMS stars}

The physical processes taking place during the formation of stars are extremely important for the next stages of their evolution. 
The final mass of stars accumulates gradually, and accretion from the circumstellar disk can take several million years. 
Meanwhile, the disk is replenished by circumstellar matter left over from the formation of the protostar's core. 
About half the mass of newly formed stars is accreted by the time they become observable in the optical region as PMS stars. 
This process takes place during episodes of enhanced accretion, when the accretion rate from the circumstellar disk increases by 2-3 orders of magnitude over hundreds of years (Hartmann $\&$ Kenyon 1996). 
Circumstellar disc masses are typically less than 1$\%$ of the stellar mass, and the disc cannot replenish rapidly with new portions of circumstellar matter. 
Therefore, outbursts resulting from increased accretion probably recur every few thousand years.

In the era of all-sky photometric monitoring conducted at multiple wavelengths, the literature is replete with reports of recorded objects with large-amplitude photometric variability. 
Automated techniques have been developed to identify bursts of young stellar objects in real time, using machine learning efforts to distinguish them from other similar events. 
But due to the wide variety of variability patterns in PMS stars, these objects do not have distinct light curve patterns that are similar from object to object.
Therefore, many variable phenomena may turn out to be unclassified cases. 
The reason is that different physical mechanisms can cause variability with similar amplitudes and time scales.
High-resolution spectra are needed to probe the physical state of the circumstellar disc and accretion processes.  
They allow us to distinguish, for example, between FUor outbursts and other types of variability of similar amplitudes (Semkov $\&$ Peneva 2011). 

In the optical and near-infrared ranges, the spectra of young stars provide the main information about the temperature of the gas envelope. 
For PMS stars that have low accretion rates, the optical and infrared spectra typically show a stellar photosphere in absorption. 
As the accretion rate increases, the star's spectrum is dominated by emission components, so it becomes impossible to separate the photospheric absorption component. 
In extremely strong outbursts, with very high accretion rates, implying a very high luminosity of the star, the observed spectrum becomes that of the hot inner disk, which is in absorption, i.e. spectrum of a hot supergiant.

Another main research method that complements spectral observations is interferometry in the near and mid-infrared regions, which allows very precisely to determine the parameters of circumstellar disks. This technique can help to reveal the relationship of the structure of the inner part of the disc with the reasons that give rise to the outbursts. 
Such photometric and spectral monitoring of the youngest protostars, which show signs of enhanced accretion, is not yet possible, and this complicates the study of the physics of accretion. 
The possibility of spectral observations is improving with the development of observational methods, but the necessary information about the most obscured or deeply embedded young stellar objects is still very insufficient.
 
A third very important method of studying PMS stars is the study of archival spectral and photometric data. 
Studying the history of variable objects can contribute significantly to explaining the causes of various forms of variability, such as outbursts, eclipses, or periodic phenomena of variability. 
It is also important to study the variability over the entire available range of the electromagnetic spectrum and to look for correlation between the processes in the individual areas.

\section{FU Ori and EX Lupi type of outbursts}

The registered objects that have shown eruptions of the FUor or EXor type are relatively small in number. 
In many cases, it is still disputed exactly what type of outburst was observed. 
And there are also cases where an object is reported as a FUor or EXor, based on just a few observations, and then turns out to actually be another type of object altogether, such as a long-period variable star or an eclipsing system. 
The total number of known PMS stars in which large-amplitude outbursts have been observed is several dozen.
There are several publications in which an attempt has been made to present a list of objects classified as FUor stars. 
For example, the paper by Reipurth $\&$ Aspin (2010) presents a list of 10 FUor objects and 10 FUor-like objects that have some FUor spectral and photometric characteristics but outburst have not been observed. 
In the paper of Audard et al. (2014), a list containing 10 FUor objects and 16 FUor-like objects is presented. 
And in the paper by Connelley $\&$ Reipurth (2018), a list of 14 FUors and 10 FUor-like objects is presented. 
A comparison of these and similar papers shows that there is no consensus on the classification of FUor type of objects. 
Some objects in one publication are classified as FUors and in another as FUor-like objects.

Any case where a large-amplitude outburst of a PMS star is observed raises the question: FUor or EXor? (Ibryamov $\&$ Semkov 2021).
The main differences between these two types are the spectral (the presence or absence of certain spectral lines, their profiles and intensity) and photometric properties (the duration and amplitude of the burst and the shape of the light curve). 
The spectral variability is explained by the different sizes of the star and circumstellar envelope: the absorption regions are significantly larger than the star itself, but the expanding circumstellar envelope is also significantly larger than the absorption regions around the disc.

Strong photometric variability at maximum brightness is typical of EXors but not for FUors. 
However, similar short-term dips in brightness have been recorded for several FUor objects. 
One of the most famous such events was the minimum in the light curve of V1515 Cyg in 1980, a sharp decrease in brightness of about 1.5 mag. ($B$ band) for several months. 
This minimum in the brightness of V1515 Cyg is explained by obscuration from dust material ejected from the star (Kenyon et al. 1991). 
A short dip in brightness (0.4 mag. in $I$ band) in the light curve of the FUor object V733 Cep was observed in 2009 (Peneva et al. 2010). 
Evidence for a strong brightness variability ($\Delta$V=1.2 mag.) during the time of maximum light during the period from 1986 to 1992 is documented in our photometric study of another FUor object V1735 Cyg (Peneva et al. 2009).

The large-amplitude variability of FUors may result from the superposition of the two processes, variable accretion rate and time variable extinction (Semkov $\&$ Peneva 2012). 
In recent papers, such a scenario has been used to explain the brightness variability of two objects with characteristics intermediate between FUors and EXors V1647 Ori (Aspin et al. 2009, Aspin 2011) and V2492 Cyg (Hillenbrand et al. 2013, K\'{o}sp\'{a}l et al. 2013). 
The time variable extinction appears to be characteristic not only of some Herbig Ae/Be stars (UXor type variables), but also a common phenomenon during the evolution of all types of PMS stars. 
In the case of FUor object V582 Aur (Semkov et al. 2013), we have direct evidence from multicolor photometry indicating the presence of dust around the star.

One possible cause of the variable accretion could be fragmentation of the circumstellar disk. 
Since FUor phenomenon is likely to be repeatable, it is assumed that almost every protostar goes through several episodes of enhanced accretion, in which the initial mass of the star increases. 
Stamatellos et al. (2012) suggest that periods of episodic increased accretion may have triggered the initial fragmentation of the circumstellar disc. 
In the early stages of stellar evolution, disk fragmentation is not possible and accretion onto the stellar surface is assumed to proceed at a constant rate. 
After several episodic increases in the accretion rate, the circumstellar disk gradually fragments and thus prevents new outbursts of FUor type, or at least changes their parameters. 
Therefore, it can be assumed that the outbursts of FUors, during different periods of the stellar evolution, can vary in amplitude, duration and shape of the light curve, caused by the different fragmentation state of the disk. 
Strong accretion bursts can also trigger a mechanism to form planets of various masses inside the circumstellar disk.

The photometric data we collected confirm the diversity in the shape and type of light curves of FUor objects. 
Our knowledge of the processes occurring during FUor type of outbursts is still incomplete, and more data need to be collected from regular photometric and spectral monitoring. 
Attempts to classify FUor objects based on their photometric properties have so far been unsuccessful due to the small number of objects of this type. 
A comparison of the light curves of known FUors show that they are very different from each other and very rarely repeated. 
Even the first three 'classical' FUors (FU Ori, V1057 Cyg and V1515 Cyg) show very different rates of brightness increase and decrease (Clarke et al. 2005). 
The variety of light curves increases even more with the number of well-studied FUors, to which our work also contributes.

As a rule, the light curves of FUors are usually asymmetric, with a rapid increase and gradual decrease in brightness. 
Some objects show a very rapid increase in brightness over several months or a year, such as FU Ori, V1057 Cyg and V2493 Cyg (Semkov et al. 2010, 2012, 2021b, Clarke et al. 2005, Kopatskaya et al. 2013). But in other cases, such as V1515 Cyg, V1735 Cyg, V733 Cep and V900 Mon, the brightness increase can last for several years and even reach 20-30 years (Clarke et al. 2005, Peneva et al. 2009, 2010, Semkov et al. 2021a). 
Usually, the decline in brightness takes several decades and is very likely to reach a century. 
But there are objects where a relatively rapid decrease in brightness is observed. 
For example, V960 Mon, where the brightness decreases by 2 mag. in $V$ band over a period of about five years (Takagi et al. 2020). 
In our study of the FUor object V582 Aur, we have observed three deep minimums of the star's brightness of about 3 mag. ($R$ band), separated by periods of about five years (Semkov et al. 2013, \'{A}brah\'{a}m et al. 2018).

But there are also objects that for long periods of time, practically do not change their brightness, as in the cases of V1735 Cyg (Peneva et al. 2009) and Parsamian 21 (Semkov $\&$ Peneva 2010). 
In this respect, the light curve of the FUor object V733 Cep is unique, with its roughly symmetrical shape (Peneva et al. 2010). 
This variety of photometric properties strongly supports the assumption that FUor objects are not a homogeneous group and that the causes of this phenomenon may be several mechanisms of a different nature (Vorobyov et al. 2021).

\section{Eclipses of UX Ori type}

The results of our observations of PMS stars strongly suggest that UXor type of variability is a widespread phenomenon (Semkov et al. 2015, 2019). 
It is typical not only of Herbig Ae/Be stars and T Tauri stars from early spectral types, but also of T Tauri stars of late spectral types and stars with relatively lower masses (Semkov et al. 2017). 
This result can be explained by the low efficiency of star formation where most of the mass of molecular clouds does not participate in star formation. 
This matter remains in the vicinity of the protostars, in a number of cases forming inhomogeneous dust clouds moving in orbit around them.
Light curves over long periods of observations provide strong evidence that the deep minimums in the brightness of GM Cep is caused by obscuration of the star by circumstellar dust structures (Semkov $\&$ Peneva 2012). 

The inhomogeneity of the dust clouds may indicate an advanced evolution of the protoplanetary disk in the transition from micron-sized dust particles to the formation of kilometer-sized planetesimals (Chen et al. 2012). 
Accretion, combined with viscous light scattering, heats the circumstellar disk of the young stellar object. 
As accretion slows and the size of the dust particles grows, the disk becomes passive, in the sense that the dust absorbs starlight, heats up and re-radiates in the infrared region.
It can be argued that both the light transit time and the observed obscurations by dust particles provide a reasonable model of the absorbing medium. 
Around the stars there is a region with dimensions of several tens of astronomical units, forming a circumstellar disk, which is in the form of a ring or a spiral structure. 
This region is tens of AU from the star and consists mostly of particles around 10 $\mu$m or larger.

\section{Weak line T Tauri stars, classical T Tauri stars and Herbig Ae/Be stars}

The most common variability in T Tauri and Herbig Ae/Be stars is periodic or aperiodic variability with small amplitudes. 
In many cases, more than one type of variability can be observed on the same star.
Periodic variability in most cases is explained by the presence of spots of reduced temperature, which are located on the star surface. 
The presence of such spots, by analogy with the Sun, is the result of magnetic activity.
By examining the amplitude of the observed variability, we can determine the intensity of the magnetic field and the location of the spots. 
The results obtained for some of the objects in this dissertation show that such cool spots can persist for several years, as we do not observe a change in ephemeris, or a large change in amplitude, between individual rotation periods (Poljan\v{c}i\'{c} Beljan et al. 2014, Ibryamov et al. 2015). 
 
Large-amplitude periodicity in PMS stars is usually observed in very few cases and is usually associated with the presence of hot spots due to accretion from the circumstellar disk. 
Usually, instabilities in the disc lead to the formation of a flow directed towards the surface of the star and oriented along the magnetic field lines. 
Spots on the stellar surface that have a decreased or increased temperature can migrate along the stellar coordinates and change their area and temperature. 
This process is demonstrated by examples of several T Tauri stars with weak lines in the IC 348 region, for which periodicity is available, but the light curves for different rotation periods have a different shape (Nordhagen et al. 2016). 
Based on our observations, we would not be able to register such a change in the shape of the light curve, and the programs we used would show a lack of periodicity. 
The reason is that our data do not have a dense coverage of several hours of observations on consecutive nights, but are scattered over long periods of time.

{\it Acknowledgments:} 
This work was partly supported by the Bulgarian Scientific Research Fund of the Ministry of Education and Science under the grants DN 08-1/2016 and DN 18-13/2017.
The authors thank the Director of Skinakas Observatory Prof. I. Papamastorakis and Prof. I. Papadakis for the award of telescope time. 
This research has made use of the NASA's Astrophysics Data System Abstract Service, the SIMBAD database and the VizieR catalogue access tool, operated at CDS, Strasbourg, France.

\end{document}